# Nonpolar *m*-plane GaN/AlGaN heterostructures with intersubband transitions in the 5–10 THz band

C. B. Lim,[1,2] A. Ajay,[1,2] C. Bougerol,[1,3] B. Haas,[1,2] J. Schörmann,[4] M. Beeler,[1,2] J. Lähnemann,[1,2] M. Eickhoff,[4] and E. Monroy[1,2]

[1] University Grenoble-Alpes, 38000 Grenoble, France
[2] CEA, INAC-SP2M, 17 av. des Martyrs, 38000 Grenoble, France
[3] CNRS, Institut Néel, 25 av. des Martyrs, 38000 Grenoble, France
[4] I. Physikalisches Institut, Justus-Liebig-Universität Gießen, Heinrich-Buff-Ring 16, 35392 Gießen, Germany

## Abstract

This paper assesses intersubband transitions in the 1–10 THz frequency range in nonpolar *m*-plane GaN/AlGaN multi-quantum-wells deposited on free-standing semi-insulating GaN substrates. The quantum wells were designed to contain two confined electronic levels, decoupled from the neighboring wells. Structural analysis reveals flat and regular quantum wells in the two perpendicular in-plane directions, with high-angle annular dark-field scanning transmission microscopy images showing inhomogeneities of the Al composition in the barriers along the growth axis. We do not observe extended structural defects (stacking faults or dislocations) introduced by the epitaxial process. Low-temperature intersubband absorption from 1.5 to 9 THz (6.3 to 37.4 meV) is demonstrated, covering most of the 7–10 THz band forbidden to GaAs-based technologies.

# 1. Introduction

The development of compact solid-state THz sources is driven by its numerous applications in biological and medical sciences, industrial and pharmaceutical quality control, security screening and communication. GaAs-based quantum cascade lasers (QCLs) have proven their potential as emitters in the 0.85–5 THz range [1,2], although the low maximum operation temperature remains a hurdle for their commercialization. There are two major processes that cause the degradation of population inversion in THz QCLs at higher temperatures [3]:

(i) Backfilling of the lower radiative state with electrons from the heavily populated injector, which occurs either by thermal excitation, or by reabsorption of non-equilibrium longitudinal optical (LO) phonons.

(ii) Thermally activated LO-phonon scattering, as electrons in the upper radiative state acquire sufficient in-plane kinetic energy to emit an LO-phonon and relax nonradiatively to the lower radiative state.

Both of these mechanisms sensitively depend on the electron gas temperature, which is 50-100 K higher than the lattice temperature during device operation. Thus, the GaAs LO phonon energy of 36 meV is a relatively low barrier for these thermally-activated phenomena. GaN, with an LO phonon energy of 92 meV (22.2 THz), opens prospects for room temperature THz lasers [4,5]. Furthermore, phonon absorption in GaAs hinders the extension of these technologies towards the 5–10 THz range, and the blue shift of the Reststrahlen band in GaN open the possibility of other intersubband (ISB) devices covering this 5-10 THz GaAs "forbidden" band.

So far, studies on ISB transitions in group-III-nitride multi-quantum-wells (MQWs) have mostly focused on polar *c*-plane structures [6]. However, in this crystallographic orientation, optical properties are affected by a polarization-induced internal electric field, which renders ISB transition energies more sensitive to the strain state of the quantum wells (QWs) [7], and hampers the extension of ISB transitions towards far-infrared wavelengths [8]. Although the electric field can be partially compensated by the implementation of multi-layer QW designs [9–12], it is still a major hurdle for device design. The use of nonpolar crystallographic orientations leads to GaN/AlGaN heterostructures without internal electric field, which facilitates the device design while maintaining the benefits of GaN.

We have recently reported that the $(1\bar{1}00)$ *m* plane is the most promising nonpolar crystallographic orientation for ISB applications, based on comparative results with the $(11\bar{2}0)$ *a* plane [13]. Mid-infrared ISB absorption in the 4.0 to 5.8 µm (310 to 214 meV) range has been observed on *m*-plane GaN/AlGaN multi-quantum-wells (MQWs) [13,14] and photodetection at 7.5 and 9.3 µm (165 and 133 meV, respectively) has been demonstrated at 14 K using *m*-InGaN/(Al)GaN MQWs [15]. Recently, low-temperature ($T = 9$ K) ISB absorption has been shown in the 3.77 to 6.31 THz range (15.6 to 26.1 meV) using *m*-GaN/AlGaN MQWs [16]. However, ISB transitions in the 7–10 THz band, inaccessible to As-based technologies, have not been reported in nonpolar nitrides yet. In this work, we investigate ISB transitions in nonpolar *m*-plane GaN/AlGaN MQWs in the 1–10 THz band.

## 2. Sample design and experimental section

A series of *m*-plane GaN/AlGaN MQW structures were designed using the Nextnano$^3$ 8×8 k.p self-consistent Schrödinger-Poisson solver [17] with the material parameters described in Ref. [7]. A cell consisting of three QWs with periodic boundary conditions was used to model the MQW structure. Using this setup, the electronic decoupling of the QWs can be confirmed. The thickness of the barriers was fixed at approximately 20 nm to avoid coupling between QWs, and the QW widths were chosen to display ISB absorption between the ground conduction band level ($e_1$) and the first excited electronic level ($e_2$) in the 4.8–8 THz (19.7–33 meV) range. Additionally, the Al content of the barriers was varied to keep the two lowermost electronic levels confined in the QW and the third electronic level close to the continuum. Table 1 summarizes the *m*-plane MQW architectures considered in this paper, and figure 1 shows the calculated band diagram corresponding to sample S2 in table 1.

Structures consisting of 40 periods of GaN/AlGaN MQWs were deposited on free-standing semi-insulating *m*-GaN platelets sliced from (0001)-oriented GaN boules synthesized by hydride vapor phase epitaxy (resistivity $>10^6$ Ω.cm, dislocation density $<5\times10^6$ cm$^{-2}$). The MQW structures were capped with a 50 nm AlGaN layer with the same Al content as used for the barriers. The samples were grown by plasma-assisted molecular-beam epitaxy (PAMBE) at a substrate temperature of 720 °C and with a nitrogen-limited growth rate of 0.5 ML/s (≈ 450 nm/h). Growth was performed under the optimum conditions for *c*-GaN, i.e. slightly Ga-rich conditions [7,18,19], which are known to be compatible with *m*-plane growth [13,20]. The GaN wells were homogeneously doped with silicon at a concentration of $3\times10^{18}$ cm$^{-3}$.

The samples were analyzed by High-Angle Annular Dark-Field Scanning Transmission Electron Microscopy (HAADF-STEM) and High-Resolution Transmission Electron Microscopy (HR-TEM) performed in an FEI Titan Ultimate microscope operated at 200 kV.

High-resolution x-ray diffraction (HR-XRD) measurements were done using a PANalytical X'Pert PRO MRD system. Experimental measurements were compared with simulations using the X'Pert Epitaxy software from PANalytical.

Fourier transform infrared spectroscopy (FTIR) was performed at 5 K using a Bruker V70v spectrometer equipped with a mercury lamp, a Si beam splitter, and a helium-cooled Si bolometer. Two pieces of each sample, each with a width of 4 mm and a thickness of 320 µm, were polished at 45° to form multipass waveguides. The pieces were placed face-to-face on the cold finger of a liquid helium-cooled cryostat, and were compared to pieces of *m*-GaN substrates prepared in the same fashion. All of the samples were tested in transmission mode using a far-infrared polarizer to discern between the transverse-electric (TE) and transverse-magnetic (TM) polarized light.

## 3. Results and discussion

To evaluate their structural quality, the samples were analyzed by HAADF-STEM and HR-TEM. Figures 2(a) and (b) show cross-section HAADF-STEM images of sample S2, viewed along (a) <0001> and (b) <1$\bar{1}$20>. Layers with dark and bright contrast correspond to the AlGaN barriers and GaN QWs, respectively. The images show the flatness of the QW interfaces along the two perpendicular directions. Furthermore, no dislocations or stacking faults appear in the epitaxial layers, which was further confirmed by HR-TEM images. Figure 3(a) presents a HAADF-STEM view of two QWs in the middle of the same sample. The AlGaN barriers present an inhomogeneous contrast with dark lines parallel to the QW interfaces. This contrast is assigned to alloy inhomogeneity along the growth axis, as confirmed by the absence of extended defects in high-resolutions images [see figure 3(b)]. From the intensity profile extracted from figure 3(a), the alloy fluctuations in the barriers can reach ±30% of the average concentration, which has been further confirmed by energy-dispersive x-ray spectroscopy (not shown).

The periodicity of the MQWs was assessed by HR-XRD. Figure 4 presents ω-2θ scans along the (3$\bar{3}$00) reflection of samples S1 and S4, together with simulations, assuming that the quantum structures are fully strained on the GaN substrate. Table 1 summarizes the MQW periods extracted from the inter-satellite distance in the HR-XRD measurements. The full width at half maximum (FWHM) of the rocking curves was measured for the substrate and the MQW zero-order ($3\bar{3}00$) reflection with $\phi = 0°$ and $\phi = 90°$ ($\Delta\omega_c$ and $\Delta\omega_a$, respectively). These broadening values provide information about the sample mosaicity in the *c* and *a* directions, respectively. Both $\Delta\omega_c$ and $\Delta\omega_a$ remain in the 30±8 arcsec range for all the MQWs, and these values are similar to those measured for the substrate reflections. This result is consistent with the absence of epitaxially-generated defects in the transmission electron microscopy images.

To probe the ISB absorption of the MQWs, characterization by FTIR spectroscopy was performed at 5 K. Using sample S2 as an example, figure 5(a) illustrates typical TE- and TM-polarized THz transmission measurements, and figure 5(b) compares the same transmission spectrum for TM-polarized light with that of the substrate. The apparent noise superimposed on all the spectra is an oscillation with nearly-regular periodicity in energy [see magnified view in the inset of figure 5(a)], which is assigned to a Fabry-Pérot interference. Using the refractive index of GaN in the far-infrared range from Ref. [21], the cavity length associated to the interference is ≈ 350 µm, corresponding to the overall thickness of the samples. The transmission spectra for TE-polarized light present additional Fabry-Pérot oscillations associated to the MQW layers. In contrast, the transmission spectrum for TM-polarized light exhibits a broad dip, in the 3-8 THz range in the case of sample S2, which is assigned to ISB absorption, following the polarization selection rule.

Figure 6 presents the normalized absorbance of samples S1 to S4 for TM-polarized light, extracted from the transmission measurements. TM-polarized absorption is observed over a broad spectral window of normalized bandwidth close to 1, whose extreme values and central energy are summarized in table 1. The central energy decreasing from 27.1 to 20.9 meV (6.5 to 5 THz) as the QW width increases is consistent with the trend of the simulations. The broad absorption bands are consistent with the doping density in the QWs of around $n_S = 3\times10^{12}$ cm$^{-2}$, i.e. at least three times higher than in Ref. [16], as illustrated in the inset of figure 5(b). The samples show ISB absorption ranging from 6.3 to 37.4 meV (1.5 to 9 THz), providing experimental evidence that ISB transitions in GaN MQWs can cover the THz spectral range forbidden to GaAs.

## 4. Conclusions

In summary, we have designed a series of nonpolar *m*-plane GaN/AlGaN MQWs by varying the QW thicknesses and Al compositions to separate the two confined electronic levels by 20–33 meV (corresponding to 4.8–8 THz transitions), and decouple these transitions from the neighboring wells. The samples were grown by PAMBE on free-standing semi-insulating GaN substrates, and the structural analysis showed MQWs composed of flat and regular layers in the two perpendicular in-plane directions, with inhomogeneities of the Al composition in the barriers along the growth axis. Extended defects introduced by the epitaxial process, such as stacking faults or dislocations, were not observed. Optically, the structures display low-temperature ISB absorption with a normalized bandwidth close to 1, which is attributed to the high doping level. This absorption occurs in the 6.3 to 37.4 meV (1.5 to 9 THz) range, providing an experimental demonstration of the possibility for GaN to cover a large part of the 7–10 THz band forbidden to GaAs-based technologies.

## Acknowledgements

Thanks are due to N. Mollard for sample preparation by focused ion beam at the NanoCharacterization Platform (PFNC) in CEA-Minatec Grenoble. This work is supported by the EU ERC-StG "TeraGaN" (#278428) project. A.A. acknowledges financial support by the French National Research Agency via the GaNEX program (ANR-11-LABX-0014).

## Tables

**Table 1.** Structural and optical characteristics of the *m*-plane GaN/AlGaN MQWs: QW and barrier thickness ($t_{QW}$ and $t_B$, respectively); Al composition of the barriers ($x_B$); MQW period measured by HR-XRD; FWHM of the ω-scan of the $(3\bar{3}00)$ x-ray reflection of the MQWs and of the GaN substrate, measured in the *c* and *a* directions ($\Delta\omega_c$ and $\Delta\omega_a$, respectively); simulated ISB transition energy; measured ISB transition energy window and central energy.

| Sample | $t_{QW}$ (nm) | $t_B$ (nm) | $x_B$ (%) | Period (nm) | ω-scan FWHM MQW (arcsec) | ω-scan FWHM GaN (arcsec) | Simulated ISB transition energy (meV) | Measured ISB transition energy window (meV) | Measured ISB transition central energy (meV) |
|---|---|---|---|---|---|---|---|---|---|
| S1 | 9.5 | 21.7 | 8 | 31.2 | $\Delta\omega_c = 35$<br>$\Delta\omega_a = 34$ | $\Delta\omega_c = 34$<br>$\Delta\omega_a = 39$ | 33 | [13.4 – 37.4] | 25.4 |
| S2 | 10.0 | 18.5 | 7.5 | 28.5 | $\Delta\omega_c = 38$<br>$\Delta\omega_a = 35$ | $\Delta\omega_c = 35$<br>$\Delta\omega_a = 36$ | 30.5 | [18.1 – 36.1] | 27.1 |
| S3 | 10.4 | 21.2 | 7.5 | 31.6 | $\Delta\omega_c = 22$<br>$\Delta\omega_a = 31$ | $\Delta\omega_c = 24$<br>$\Delta\omega_a = 39$ | 30.3 | [6.3 – 36.3] | 21.3 |
| S4 | 12.9 | 21.1 | 6 | 34.1 | $\Delta\omega_c = 38$<br>$\Delta\omega_a = 28$ | $\Delta\omega_c = 42$<br>$\Delta\omega_a = 29$ | 19.7 | [7.9 – 33.9] | 20.9 |

# Figure captions

**Figure 1.** Conduction band diagram with the three first electronic levels and their squared wavefunctions of a QW in the center of the active region of sample S2.

**Figure 2.** Cross-section HAADF-STEM images of sample S2 viewed (a) along <0001>, and (b) along <1$\bar{1}$20>. Layers with dark and bright contrast correspond to the AlGaN barriers and GaN QWs, respectively.

**Figure 3.** Layers with dark and bright contrast correspond to the AlGaN barriers and GaN QWs, respectively. (a) Cross-section HAADF-STEM image of sample S2 viewed along <1$\bar{1}$20> and intensity profile along <0001>. (b) High-resolution HAADF-STEM image of the barrier/QW interface showing that the variations of contrast in the image are not associated to structural defects.

**Figure 4.** HR-XRD ω-2θ scans of the (3$\bar{3}$00) reflection of samples S1 and S4. Simulations assume the quantum structures fully strained on the GaN substrate.

**Figure 5.** (a) Transmission of TM and TE-polarized light of sample S2 measured in the THz range. Inset: Magnification of the transmission spectra showing Fabry-Pérot oscillations. (b) Transmission of TM-polarized light of sample S2 and of the substrate, measured in the THz range. Inset: Normalized broadening of the absorbance (energy broadening divided by the central energy) as a function of the doping density in the MQWs. Hollow square symbols are extracted from Ref. [16], and the full round symbol corresponds to this work.

**Figure 6.** Normalized absorbance of TM-polarized light for samples S1 to S4 in the THz range. Data are vertically shifted for clarity. The striped and shadowed areas represent the phonon absorption bands of GaAs and GaN, respectively.

**Figure 1**

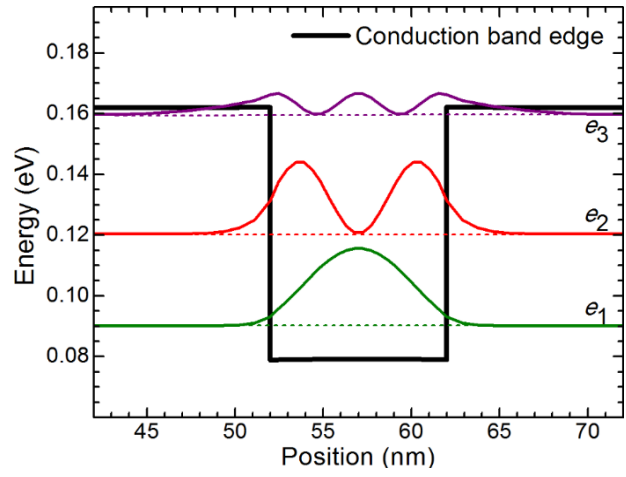

**Figure 2**

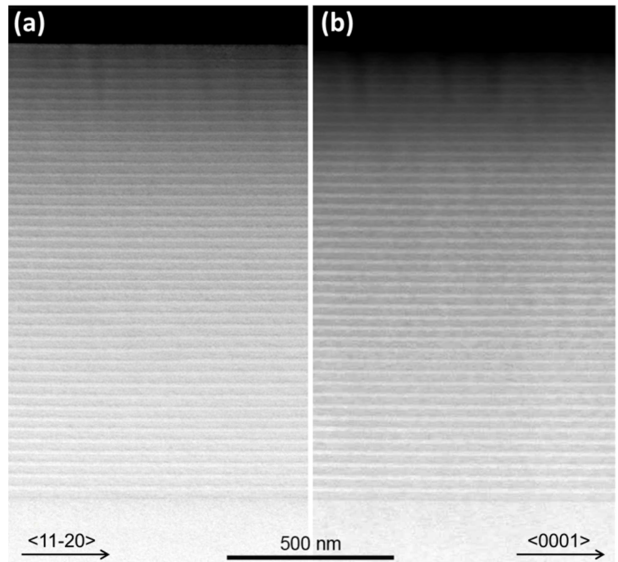

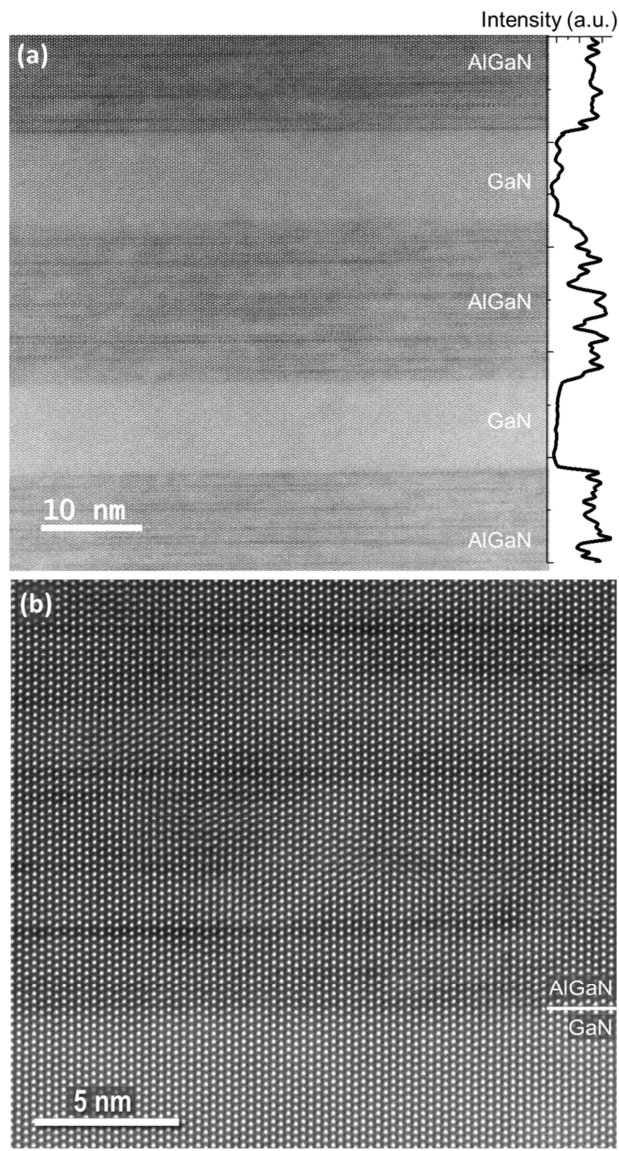

**Figure 4**

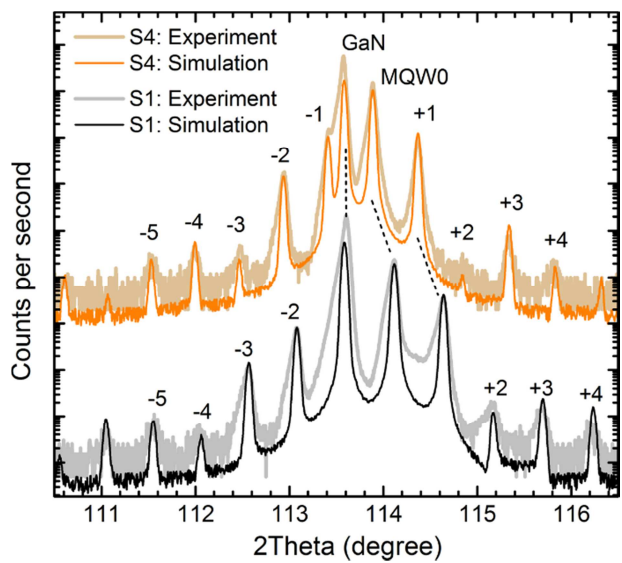

**Figure 5**

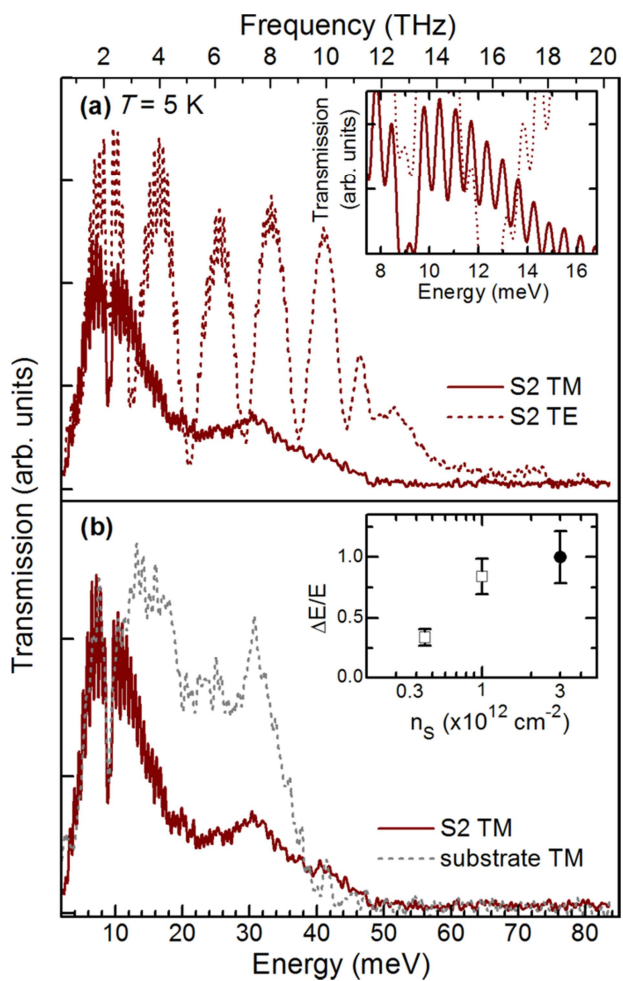

**Figure 6**

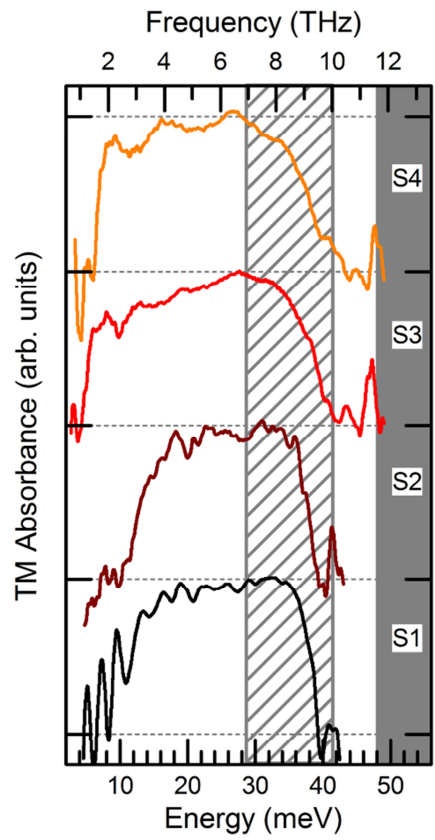